\documentclass[11pt]{article}
\usepackage{amsfonts}
\begin{document}
\def\bbox#1{{\mathbf{#1}}}
\def\text#1{{\mathrm{#1}}}
\newcommand{\dbar}{\bar{\partial}}
\newcommand{\wt}{\widetilde}
\newcommand{\be}{\begin{equation}}
\newcommand{\ee}{\end{equation}}
\newcommand{\bea}{\begin{eqnarray}}
\newcommand{\eea}{\end{eqnarray}}
\newcommand{\beaa}{\begin{eqnarray*}}
\newcommand{\eeaa}{\end{eqnarray*}}
\newcommand{\qd}{\quad}
\newcommand{\qqd}{\qquad}
\newcommand{\npb}{\nopagebreak[1]}
\newcommand{\nn}{\nonumber}
\renewcommand{\theequation}{\thesection.\arabic{equation}}
\title
{Quasi-classical $\dbar$-method: Generating equations for
dispersionless integrable hierarchies}
\author{
L.V. Bogdanov\thanks
{L.D. Landau ITP, Kosygin str. 2,
Moscow 117940, Russia}, 
B.G. Konopelchenko\thanks
{Dipartimento di Fisica dell' Universit\`a
and Sezione INFN, 73100 Lecce, Italy}
~and 
L. Martinez Alonso\thanks
{Departamento de Fisica Teorica II Universidad Complutense,
Madrid, Spain}
}
\date{}
\maketitle
\begin{abstract}
The quasi-classical $\bar{\partial}$-dressing method
is used to derive compact generating equations
for dispersionless hierarchies. Dispersionless Ka\-dom\-tsev-Petviashvili
(KP) and two-dimensional Toda lattice (2DTL) hierarchies are considered
as illustrative examples.
\end{abstract}
\section{Introduction}
\setcounter{equation}{0}
Dispersionless integrable equations and corresponding hierarchies 
have attracted a considerable interest during the last two decades.
Their study is of great importance since dispersionless 
integrable hierarchies constitute an essential part
of the general theory of integrable systems (see e.g. \cite{KM77}-\cite{13})
and they arise in the analysis of many problems in physics,
mathematics and applied mathematics (see e.g. \cite{14}-\cite{DMT}).

Recently it was shown that dispersionless integrable hierarchies
are ame\-nable to the quasi-classical $\dbar$-dressing method
\cite{KMR,KM1,KM2}.This approach provides us with a simple and elegant method 
of constructing and solving dispersionless integrable hierarchies.
Moreover, it establishes a connection between these
hierarchies and the theory of
quasi-conformal mappings on the plane \cite{KM1}.

In the present paper we use the quasi-classical $\dbar$-dressing 
method in order to derive compact generating equations for
dispersionless integrable hierarchies. These equations, in particular,
imply the existence of $\tau$-functions and, in a very simple way,
provide us with the corresponding dispersionless addition formulae
(Hirota equations). The dispersionless KP (dKP) and 2DTL (d2DTL)
hierarchies are considered as illustrative examples of the general
approach.
\section{Quasi-classical $\dbar$-method}
\setcounter{equation}{0}
The quasi-classical $\dbar$-dressing method proposed in \cite{KMR,KM1,KM2}
is based on the quasi-classical $\dbar$-problem
\be
S_{\bar z}=W(z, \bar z, S_z),
\label{ddbar}
\ee
where $z\in\mathbb{C}$, bar means complex conjugation,
$$
S_z=\frac{\partial S(z,\bar{z})}{\partial z}
$$
and $W$ (quasi-classical $\dbar$-data) is an analytic function
of $S_z$. We are looking for solutions of equation
(\ref{ddbar}) in the form
$$
S=S_0+\wt S,
$$
where ${S_0}_{\bar z}=0$ for $z\in G$ ($G$ is certain domain in
$\mathbb{C}$) and $\wt S_{\bar z}=0$ for $z\in\mathbb{C}\setminus G$.
Parameterizing a function $S_0$ which is analytic in $G$
by certain (infinite) set of parameters (times),
i.e., $S_0=S_0(z; t_1,t_2,\dots)$, one thus has solutions of equation 
(\ref{ddbar}) depending on these parameters:
$S=S(z,\bar z;t_1,t_2,\dots)$.
A standard way to describe such a dependence is provided
by partial differential equations for $S$ with $t_1,t_2,\dots$
as independent variables.

To derive these equations, one notices that equation
(\ref{ddbar}) implies that
\be
\left( \frac{\partial S}{\partial t_i}\right)_{\bar z}=
W'(z,\bar z; S_z)\left(\frac{\partial S}{\partial t_i}\right)_z,
\quad i=1,2,3,\dots,
\label{ddbarl}
\ee
where
$$
W'(z,\bar z; \xi)=\frac{\partial W(z,\bar z; \xi)}
{\partial \xi}.
$$
The Beltrami equation on the plane
$$
f_{\bar z}=\mu f_z
$$
has a number of remarkable properties
(see e.g. \cite{Vekua}).
Two of them are crucial for the quasi-classical $\dbar$-method.
They are:
\\
1. if $f_1$ and $f_2$ are solutions of the Beltrami equation 
then $F(f_1,f_2)$ where $F$ is an arbitrary
differentiable function is a solution too;
\\
2. if $f$ is the solution of the Beltrami equation which is bounded in
$\mathbb{C}$ and $f\rightarrow 0$ as $z\rightarrow z_0$ 
($z_0\in \mathbb{C}\cup\infty$),
then (under certain mild conditions) $f=0$.

These two properties imply that, first,
$F(\frac{\partial S}{\partial t_1},\frac{\partial S}{\partial t_2},\dots)$,
where $F$ is an arbitrary differentiable function, obeys the Beltrami equation
$F_{\bar z}=\mu F_z$ and second, if $F$ is bounded as a function of
$z$ and $F\rightarrow 0$ as $z\rightarrow z_0$, then $F\equiv0$.
Thus such functions provide us with equations \cite{KM2}
\be
F_i\left(
\frac{\partial S}{\partial t_1},\frac{\partial S}{\partial t_2},\dots
\right)
=0,
\quad i=1,2,\dots
\label{DE}
\ee
These equations are our desired differential equations for dispersionless
hierarchies. The form of equations (\ref{DE}) is completely determined
by the choice of the domain $G$ and parameterization of 
$S_0(z;t_1,t_2,\dots)$.

Thus, the quasi-classical $\dbar$-dressing method consists basically
in rather elementary operations of complex analysis applied to equation
(\ref{ddbar}).
\section{dKP hierarchy}
\setcounter{equation}{0}
In the paper \cite{KM2} the quasi-classical $\dbar$-dressing method has been
used to derive dKP, dmKP and d2DTL hierarchies in their usual formulations.
Here we present the derivation of generating equations which encode different
forms of dispersionless hierarchies, and thus play a central role in their
theory.

We begin with the dKP hierarchy. In this case the domain $G$ is the unit
disk ($|z|\leq1$), $W=0$ for $|z|>1$, and $S_0=\sum_{n=1}^{\infty}z^nt_n$
\cite{KM2}. We also require $\wt S\sim \wt S_1z^{-1}+ \wt S_2z^{-2}+\dots$
as $z\rightarrow\infty$. The quantity $p=\frac{\partial S}{\partial t_1}$
is a basic homeomorphism \cite{KM1} and
$$
p=z+\frac{1}{z}\frac{\partial \wt S_1}{\partial t_1}+\dots
$$
as $z\rightarrow\infty$. Let us introduce the well-known operator
$$
D(z)=\sum_{n=1}^{\infty}\frac{1}{n}\frac{1}{z^n}\frac{\partial}{\partial t_n}.
$$
In virtue of the properties of equation
(\ref{ddbar}) and Beltrami equation (\ref{ddbarl}) the function
$D(z_1)S(z)$ ($z\in\mathbb{C}$,
$z_1\in\mathbb{C}\setminus G$) and consequently $\exp(-D(z_1)S(z))$
obey equation (\ref{ddbarl}).
Since $D(z_1)S_0(z)=-\log(1-\frac{z}{z_1})$, one has
\be
\exp(-D(z_1)S(z))=\left(1-\frac{z}{z_1}\right)\exp(-D(z_1)\wt S(z)).
\label{exp}
\ee
Thus, $\exp(-D(z_1)S(z))\sim -\frac{1}{z_1}z +O(1)$ as $z\rightarrow\infty$.
Using these properties of the functions $p(z)$ and $\exp(-D(z_1)S(z))$,
one constructs an equation of the form (\ref{DE}). Namely,
\be
p(z)+ z_1 \exp(-D(z_1)S(z))-(z_1+D(z_1)\wt S_1)=0.
\label{DE1}
\ee

On the other hand, evaluating the l.h.s. of (\ref{DE1}) at $z=z_1$
and using (\ref{exp}), one gets its equivalent form
\be
p(z)-p(z_1)+z_1\exp(-D(z_1)S(z))=0,\quad z\in\mathbb{C},
\quad z_1\in\mathbb{C}\setminus G.
\label{DE2}
\ee
It is easy to check this equation directly. Indeed, the l.h.s.
is a bounded solution of Beltrami equation (\ref{ddbarl})
having zero at $z=z_1$. Then the properties of Beltrami equation
imply that it is zero identically.

Equation (\ref{DE2}) occupies the central place in the theory of dKP hierarchy.
This fact has been already understood in \cite{YK,TT}.However, we would like to
emphasize that in equation (\ref{DE2}) $z$ is an arbitrary point 
in $\mathbb{C}$, while the corresponding equation in \cite{YK,TT}
is valid for $z\in \mathbb{C}\setminus G$ only. The fact that 
$z\in\mathbb{C}$ in equation
(\ref{DE2})  will be essential in some of further
constructions.

An immediate consequence of equation (\ref{DE2}) together with (\ref{exp})
is that for $z,z_1\in\mathbb{C}\setminus G$
$$
D(z_1)\wt S(z,\bbox{t})=D(z)\wt S(z_1,\bbox{t}).
$$
So, there exists a function $F(\bbox{t})$ such that
$$
\wt S(z,\bbox{t})=-D(z)F(\bbox{t}), \quad z\in\mathbb{C}\setminus{G}.
$$
Thus, one has (see also \cite{YK,TT,CK})
\be
p(z_1)-p(z_2)=(z_1-z_2)\exp(D(z_1)D(z_2)F),
\quad z_1,z_2\in\mathbb{C}\setminus G.
\label{add0}
\ee
Considering equation (\ref{add0}) for pairs of parameters
$(z_1,z_2)$, $(z_2,z_3)$, $(z_3,z_1)$ and summing them up,
one gets
\bea
(z_1-z_2)e^{D(z_1)D(z_2)F}+(z_2-z_3)e^{D(z_2)D(z_3)F}
+(z_3-z_1)e^{D(z_3)D(z_1)F}
\nn\\
=0,
\label{dKPadd}
\\
z_1,z_2,z_3\in\mathbb{C}\setminus G,
\nn
\eea
that is the dispersionless Fay identity \cite{TT,TT1}. 
Thus $F=\log \tau_\text{dKP}$, where $\tau_\text{dKP}$
is the $\tau$-function of the dKP hierarchy \cite{TT,TT1}.

Equation (\ref{DE2}) generates the hierarchy of Hamilton-Jacobi
type equations for the dKP hierarchy. Indeed, rewriting 
(\ref{DE2}) as 
\be
\frac{\partial S(z)}{\partial t_1}-\frac{\partial S(z_1)}
{\partial t_1}+z_1e^{-D(z_1)S(z)}=0,
\ee
and expanding its l.h.s. as $z_1\rightarrow\infty$,
one gets the equations
\bea
&&
-\frac{\partial\wt S_1(t)}{\partial t_1}-
\frac{1}{2} \frac{\partial S(z)}{\partial t_2}+
\frac{1}{2}\left(\frac{\partial S(z)}{\partial t_1}\right)^2=0,
\label{HJ}\\
&&
-\frac{\partial\wt S_2(t)}{\partial t_1}-
\frac{1}{6} \frac{\partial S(z)}{\partial t_3}
+\frac{1}{6}\frac{\partial S(z)}{\partial t_1}
\frac{\partial S(z)}{\partial t_2}
-\frac{1}{6} \left(\frac{\partial S(z)}{\partial t_1}\right)^3=0
\nn
\eea
and so on, that represents one of the equivalent forms of the known
Hamilton-Jacobi type equations for dKP hierarchy.

Equation (\ref{DE2}) allows us to derive an equation 
which generates the hierarchy of equations for 
$ S(z,\bar z;\bbox{t})$ only. In fact, acting by $D(z_k)$
on equation (\ref{DE1}) (with substitution 
$z_1\rightarrow z_i$), one gets
\bea
D(z_k)p(z)-D(z_k)D(z_i)\wt S_1+
z_iD(z_k)e^{-D(z_k)S(z)}=0,
\label{DE3}
\\
z\in\mathbb{C}, \quad z_i,z_k\in \mathbb{C}\setminus G.
\nn
\eea
Taking equation (\ref{DE3}) with all possible
pairs $z_i,z_k$ ($i,k=1,2,3$) and summing up,
one obtains the equation
\bea
\sum_{i,j,k}\epsilon_{ijk}z_i
D(z_k)\left(e^{-D(z_i)S(z,\bar z;\bbox{t})}
\right)=0,
\label{GE1}
\\
z\in\mathbb{C}, 
\quad z_1,z_2,z_3\in \mathbb{C}\setminus G,
\nn
\eea
where $\epsilon_{ijk}$ is the totally antisymmetric tensor
(with $\epsilon_{123}=1$) and summation is performed
over all $i,j,k\in{1,2,3}$. Considering equation
(\ref{GE1}) for $z_1,z_2,z_3\rightarrow\infty$
and collecting the terms of different orders
in $z_1^{-1},z_2^{-1}$ and $z_3^{-1}$, 
one gets an infinite hierarchy of differential 
equations for $S(z,\bar z;\bbox{t})$.
In the lowest nontrivial order ($z_3^{-1}z_1^{-2},\;z_3^{-1}z_2^{-2},\;\dots$)
one has
\be
\frac{\partial^2 S(z,\bar z;\bbox{t})}
{\partial t_1\partial t_3}
-\frac{3}{4}
\frac{\partial^2 S}
{\partial t_2^2}
-\frac{3}{2}
\frac{\partial^2 S}
{\partial t_1^2}
\left(
\frac{\partial S}
{\partial t_2}
-\left(
\frac{\partial S}
{\partial t_1}
\right)^2
\right)=0,
\label{E1}
\ee
which, of course, is also a consequence
of equations (\ref{HJ}).

Further, expanding the l.h.s. of (\ref{GE1})
in the Taylor series in $z^{-1}$  as 
$z\rightarrow\infty$, one gets (in the order
$z^{-1}$)
\be
\sum_{i,j,k}\epsilon_{ijk}z_i
D(z_k)\left(z_i D(z_i)\wt S_1+
\frac{1}{2}
\left(
D(z_i)\wt S_1
\right)^2
\right)=0.
\label{GE2}
\ee
This equation is the generating equation of the whole
dKP hierarchy (for $u(\bbox{t})=
-2\frac{\partial \wt S_1}{\partial t_1}$).
it is a simple check that the expansion of 
(\ref{GE2}) at $z_1,z_2,z_3\rightarrow \infty$
gives rise to the dKP equation
$$
u_{t_1t_3}=\frac{3}{2}(uu_{t_1})_{t_1}+
\frac{3}{4}u_{t_2 t_2}
$$
and higher equations.

Thus, equation (\ref{GE1}) is one of the
fundamental equations for the dKP hierarchy.
Note that since $S(z,\bar z;\bbox{t})$ defines
a quasi-conformal mapping of the domain $G$
\cite{KM1}, equation (\ref{GE1}) is also the central
equation for the integrable deformations
of quasi-conformal mappings.

Finally, we present one more equation associated
with the dKP hierarchy.
Using (\ref{DE2}), one gets
\bea
e^{D(z_1)(S(z_2)-S(z))}+
e^{D(z_2)(S(z_1)-S(z))}=1,
\label{DB1}
\\
z\in\mathbb{C},\quad z_1,z_2\in\mathbb{C}\setminus G.
\nn
\eea
Consequently,
\bea
e^{D(z_i)(S_j(\bbox{t})-S_k(\bbox{t}))}+
e^{D(z_j)(S_i(\bbox{t})-S_k(\bbox{t}))}=1,
\label{DB2}
\eea
where $i,j,k\in{1,2,3}$, $i\neq j\neq k\neq 1$,
$S_i(\bbox{t})=S(z_i, \bar z_i;\bbox{t})$,
$z_1,z_2, z_3\in\mathbb{C}\setminus G$.
This system of three equations
for three functions $S_1(\bbox{t}),
S_2(\bbox{t}),S_3(\bbox{t})$
can be considered as a dispersionless version of the
Darboux system.
\section{2DTL hierarchy}
\setcounter{equation}{0}
For the dispersionless 2DTL hierarchy the domain $G$ is an annulus
$a<|z|<b$, where $a,b$ ($a,b\in\mathbb{R}$, $a,b>0$; $b>a$) are arbitrary.
To set the quasi-classical $\dbar$-problem (\ref{ddbar}) correctly,
in general we do not need to require analyticity of the function
$S_0$ in $G$, it is enough to have analyticity of 
its derivative ${S_0}_z$.
A generic function $S_0$ 
with ${S_0}_z$ analytic in $G$ can be represented
as
$$
S_0=t\log z +\sum_{n=1}^{\infty}z^n x_n+
\sum_{n=1}^{\infty}z^{-n}y_n,
$$
where $t,x_n,y_n$ are free parameters \cite{KM2}.
We assume that $\wt S(z)\sim \sum_{n=1}^\infty \frac{S_n}{z^n}$
as $z\rightarrow\infty$ and denote $\wt S(0)=\phi$,
$G_+=\{z,|z|>b\}$, $G_-=\{z,|z|<a\}$.
The functions $p_+=\frac{\partial S}{\partial x_1}$
and $p_-=\frac{\partial S}{\partial y_1}$ have pole singularities
while $p=\frac{\partial S}{\partial t}$ has a logarithmic singularity.
The functions $\exp(-D_+(z_1)S(z))$,
$\exp(-D_-(z_2)S(z))$, $\exp(DS(z))$, where $z\in\mathbb{C}$,
$z_1\in G_+$, $z_2\in G_-$ and
$$
D_+(z)=\sum_{n=1}^\infty\frac{1}{n}\frac{1}{z^n}\frac{\partial}{\partial x_n},
\quad
D_-(z)=\sum_{n=1}^\infty\frac{1}{n}{z^n}\frac{\partial}{\partial y_n},
\quad
D=\frac{\partial}{\partial t},
$$
are solutions of the Beltrami equation and have pole singularities too.
 
Using these functions, one gets the following equations 
\bea
&&
p_+(z)-p_+(z_1)+z_1e^{-D_+(z_1)S(z)}=0,\quad z\in\mathbb{C},z_1\in G_+,
\label{TE1}
\\
&&
p_-(z)-p_-(z_2)
+\frac{1}{z_2}e^{D_-(z_2)(\phi-S(z))}=0,\quad z\in\mathbb{C},z_2\in G_-,
\label{TE2}
\\
&&
p_+(z)=e^{p(z)-D\wt S_1},\quad z\in \mathbb{C},
\label{TE3}
\\
&&
p_-(z)=e^{D\phi-p(z)},\quad z\in \mathbb{C},
\label{TE4}
\eea
Equation (\ref{TE1}) coincides with (\ref{DE2}) and gives rise to the
dKP hierarchy in variables $x_n$. In particular, one has
$$
\wt S(z_1)=-D_+(z_1)F_+,\quad z_1\in G_+
$$
as well as the corresponding dKP addition formula and generating equations.

Equation (\ref{TE2}) encodes the dmKP hierarchy.
It implies that
$$
\wt S(z_2)=\phi-D_-(z_2)F_-,\quad z_2\in G_-,
$$
the function $F_-$ obeys the dmKP addition formula
which coincides with (\ref{dKPadd}) under the substitution
$F\rightarrow F_-$, $D\rightarrow D_-$ and $z\rightarrow z^{-1}$.
It is a simple exercise to derive the corresponding generating equations.

The formulae (\ref{TE3}) and (\ref{TE4}) allow us to connect these two pieces
into the d2DTL hierarchy. Using (\ref{TE3}), one rewrites (\ref{TE4})
as
\be
e^{DS(z)}-e^{DS(z_1)}+z_1e^{-D_+(z_1)S(z)}=0,\quad z\in\mathbb{C},
z_1\in G_+,
\label{TE5}
\ee
while using the relation $p_-(z_2)=z_2^{-1}\exp(D_-(z_2)\phi)$ and
(\ref{TE4}), one transforms (\ref{TE2}) into
\be
e^{DS(z)}\left(1-e^{D_-(z_2)S(z)}\right)=
z_2e^{(D-D_-(z_2))\phi},\quad z\in\mathbb{C},z_2\in G_-.
\label{TE6}
\ee

Evaluating the l.h.s. of equation (\ref{TE5}) at $z=0$,
one gets
$$
\exp(DS(z_1))=z_1\exp (-D_+(z_1)\phi).
$$
It implies that $\phi=DF_+$.
On the other hand, evaluation of relation (\ref{TE6})
around $z_2=0$ gives
$$
\frac{\partial S(z)}{\partial y_1}
\exp(DS(z))=\exp(D\phi).
$$
Expanding the l.h.s. of this relation around $z=0$,
one gets
$$
\frac{\partial^2 F_-}{\partial y_1\partial t}=
\frac{\partial \phi}{\partial y_1}=
\frac{\partial^2 F_+}{\partial y_1\partial t}
$$
and so on. Thus, $F_+=F_-=F$.
Since $\wt S(z_1)=-D_+(z_1)F$ $(z_1\in G_+)$,
$\wt S(z_2)=D\phi-D_-(z_2)F$ $(z_2\in G_-)$,
one obtains from (\ref{TE5}) and (\ref{TE6})
the following known addition formulae 
(dispersionless Hirota equations) for the
d2DTL hierarchy (see \cite{TT1,WZ,Kost}):
\bea
\wt z_1 e^
{-DD_+(\wt z_1)F}
-z_1 e^
{-DD_+(z_1)F}
+(z_1-\wt z_1)e^
{D_+(z_1)D_+(\wt z_1)F}
=0,
\label{Tadd1}
\\
z_1,\wt z_1\in G_+,
\nn\\
1+\left(\frac{z_2}{z_1}-1\right)e^{D_-(z_2)D_+(z_1)F}
-\frac{z_2}{z_1}e^{(D_+(z_1)+D-D_-(z_2))DF}=0,
\label{Tadd2}
\\
z_1\in G_+,z_2\in G_-
\nn
\eea
plus the dKP addition formula (\ref{dKPadd}). Note that evaluating
the l.h.s. of equation (\ref{TE5}) as $z\rightarrow\infty$,
one also gets
\be
\frac{\partial^2 F}{\partial x_1\partial y_1}=
1-e^{D^2 F}.
\label{dToda}
\ee
Equations (\ref{TE5}) and (\ref{TE6}) are also the generating equations
for the hierarchy of the Hamilton-Jacobi type equations.
Indeed, expansion of the l.h.s. of equation (\ref{TE5}) for 
$z_1\rightarrow\infty$ gives
\bea
&&
\frac{\partial S(z)}{\partial x_1}-
e^{\frac{\partial S}{\partial t}}+
\frac{\partial\wt S_1}{\partial t}=0,
\\&&
\frac{\partial S}{\partial x_2}-
\left(\frac{\partial S}{\partial x_1}\right)^2
+\frac{\partial \wt S_2}{\partial t}
+\frac{1}{2}\left(\frac{\partial \wt S_1}{\partial t}\right)^2=0
\nn
\eea
and so on, while the expansion of the equations around $z_2=0$ 
provides us with the equations
\bea
&&
\frac{\partial S}{\partial y_1}-e^{\frac{\partial \phi}{\partial t}-
\frac{\partial S}{\partial t}}=0,
\\
&&
\frac{\partial S}{\partial y_2}
-\left(\frac{\partial S}{\partial y_1}\right)^2
+\frac{\partial \phi}{\partial y_1}
e^{\frac{\partial \phi}{\partial t}-
\frac{\partial S}{\partial t}}=0
\nn
\eea
plus higher equations which are equivalent to those found in \cite{KM2}.

Now let us derive generating equations for $S(z,\bar z;t,\bbox{x},\bbox{y})$
and $\phi(t,\bbox{x},\bbox{y})$. First, equation (\ref{TE5}) implies
\be
-D_+(z_1)\phi=\log
\left(
\frac{1}{z_1}e^{DS(z)}+e^{-D_+(z_1)S(z)}
\right),
\quad
z_1\in G_+,z\in\mathbb{C},
\ee
while from (\ref{TE6}) one obtains
\be
(D-D_-(z_2))\phi=\log
\left(
\frac{1}{z_2}e^{DS(z)}-\frac{1}{z_2}e^{(D-D_-(z_2))S(z)}
\right),
\quad
z_2\in G_-,z\in\mathbb{C}.
\ee
Eliminating $\phi$ from these equations, one gets
\bea
&&
(D-D_-(z_2))\log
\left(
\frac{1}{z_1}e^{DS(z)}+e^{-D_+(z_1)S(z)}
\right)
\nn
\\&&\qquad
+D_+(z_1)\log
\left(
e^{DS(z)}-e^{(D-D_-(z_2))S(z)}
\right)=0, 
\label{Tgen0}
\eea
where $z\in \mathbb{C}$, $z_1\in G_+$, $z_2\in G_-$.
Considering the limit $z_1\rightarrow\infty$,
$z_2\rightarrow0$,
one obtains in the lowest order the equation
\be
\frac{\partial^2 S}{\partial x_1\partial y_1}
+\frac{\partial S}{\partial y_1}
\frac{\partial}{\partial t}\left(e^\frac{\partial S}{\partial t}\right)=0,
\ee
which has been already found in \cite{KM2}. Higher terms in the expansion
give rise to higher equations for $S(z,\bar z;t,\bbox{x},\bbox{y})$.

To derive generating equations for $\phi$ let us rewrite equation
(\ref{Tadd2}) as
\bea
D_-(z_2)D_+(z_1)F=
\log
\left(
\frac{z_1}{z_1-z_2}
\left(
1-\frac{z_2}{z_1}e^{(D_+(z_1)+D-D_-(z_2))DF}
\right)
\right).
\nn
\eea
Applying $D$ to this equation and using the relation
$DF=\phi$, one gets
\bea
D_-(z_2)D_+(z_1)\phi=
D\log
\left(
\frac{z_1}{z_1-z_2}
\left(
1-\frac{z_2}{z_1}e^{(D_+(z_1)+D-D_-(z_2))\phi}
\right)
\right),
\label{Tgen}
\\
z_1\in G_+,z_2\in G_-.
\nn
\eea

An expansion of both sides of equation (\ref{Tgen})
as $z_1\rightarrow\infty$ and $z_2\rightarrow 0$ in the 
lowest nontrivial order ($z_1^{-1},z_2$)gives the d2DTL
equation
(see e.g. \cite{Z,TT1})
\be
\frac{\partial^2 \phi}{\partial x_1 \partial y_1}
+\frac{\partial}{\partial t}\left(
e^{\frac{\partial\phi}{\partial t}}\right)=0,
\ee
while considering higher order terms,
one gets higher equations from the d2DTL hierarchy.

The generating equations (\ref{GE1}) and (\ref{Tgen0})
represent also the compact forms of integrable deformations
of quasi-conformal mappings $S(z,\bar z;t,\bbox{x},\bbox{y})$
of the unit disk and an annulus respectively. Note that one can derive
the same formulae using the finite number of `logarithmic'
times $\xi_i$ defined by
$$
S_0(z,\bbox\xi)=\sum_{i=1}^3\xi_i\log(z-z_i)
$$
instead of infinite sets of times
$t_n$ or $x_n,y_n$.
\subsection*{Acknowledgments}
This work has been done in the framework of the grant
INTAS-99-1782. LVB was also supported in part by RFBR grants
01-01-00929 and 00-15-96007; BGK was supported in part
by the grant COFIN 2000 `Sintesi'; LMA 
was supported in part by CICYT proyecto PB98-0821.

B.G.Konopelchenko is grateful to the organizers of the 
Programme `Integrable Systems' held in the 
Isaac Newton Institute, Cambridge, UK 
for the support. L.Martinez Alonso wishes to thank the Fundacion Banco Bilbao 
Vizcaya Argentaria for its generous support during his stay at Cambridge 
University.


\begin{thebibliography}{99}

\bibitem{KM77}  
B. A. Kuperschmidt and Yu. I.  Manin, \textit{Funk. Anal. Appl.} I \textbf{11}(3),
31--42 (1977); II \textbf{17}(1), 25--37 (1978).

\bibitem{Z81}  V. E. Zakharov, \textit{Func. Anal. Priloz.} 
\textbf{14}, 89--98 (1980);
\textit{Physica} \textbf{3D}, 193--202 (1981).

\bibitem{3}  P. D. Lax and C. D. Levermore, 
\textit{Commun. Pure Appl. Math.},
\textbf{36}, 253--290, 571--593, 809--830 (1983).

\bibitem{4}  I. M. Krichever, 
\textit{Func. Anal. Priloz.}, 
\textbf{22}, 37--52 (1988).

\bibitem{YK}  Y. Kodama, 
\textit{Phys. Lett.} 
\textbf{129A}, 
223--226 (1988); \textbf{
147A}, 477--482 (1990).

\bibitem{6}  B. A. Dubrovin and S. P. Novikov, 
\textit{Russian Math. Surveys},
\textbf{44}, 35--124 (1989).

\bibitem{TT}  K. Takasaki and T. Takebe, 
\textit{Int. J. Mod. Phys. A, Suppl.}
\textbf{1B}, 889--922 (1992).

\bibitem{8}  I. M. Krichever, 
\textit{Commun. Pure Appl. Math.}, 
\textbf{47}, 437--475
(1994).

\bibitem{9} N. M.
Ercolani et al, eds., \textit{Singular limits of dispersive waves}, 
\textit{Nato Adv. Sci. Inst. Ser. B Phys.}
\textbf{320}, Plenum, New York (1994).

\bibitem{Z} V. E. Zakharov, {Dispersionless limit of
integrable systems in (2+1)-dimensions}, in \cite{9}, pp. 165--174
(1994)

\bibitem{TT1}  K. Takasasi and T. Takebe, 
\textit{Rev. Math. Phys.},
\textbf{\ 7},
743--808 (1995).

\bibitem{CK}  R. Carroll and Y. Kodama, 
\textit{J. Phys. A Math. Gen.} 
\textbf{28},
6373--6387 (1995).

\bibitem{13}  S. Jin, C. D. Levermore and D. W. McLaughlin, 
\textit{Comm. Pure and Appl. Math.}, 
\textbf{52}, 613--654 (1999).

\bibitem{14}  I. M. Krichever, 
\textit{Comm. Math. Phys.}, 
\textbf{143}, 415--429
(1992).

\bibitem{15}  B. A. Dubrovin, 
\textit{Comm. Math. Phys.} 
\textbf{145}, 195--207
(1992); 
B. A. Dubrovin, 
\textit{Nucl. Phys.} 
\textbf{B379}, 627--289 (1992);
B. A. Dubrovin and Y. Zhang, 
\textit{Comm. Math. Phys.} 
\textbf{198},
311--361 (1998).

\bibitem{16}  S. Aoyama and Y. Kodama, 
\textit{Commun. Math. Phys.} 
\textbf{182},
185--219 (1996).

\bibitem{YK97} Y. Kodama, 
\textit{SIAM J.Appl.Math.}
\textbf{59}, 2162--2192
(1999)

\bibitem{17}  Y. Gibbons and S.P. Tsarev, 
\textit{Phys. Lett.}
\textbf{258A}, 263--271
(1999).

\bibitem{MWZ00} M. Mineev-Weinstein P. B. Wiegmann and A.
Zabrodin, 
\textit{Phys. Rev. Lett.} 
\textbf{84}, 5106--5109 (2000).

\bibitem{WZ}  P. B. Wiegmann and A. Zabrodin, 
\textit{Commun. Math. Phys.}
\textbf{213}, 523--538 (2000).

\bibitem{DMT} M. Dunajski, L.J. Mason and P. Tod,
\textit{J. Geom. Phys.} 
\textbf{37}, 63--93 (2001). 

\bibitem{KMR}  B.G. Konopelchenko, L. Martinez Alonso and O. Ragnisco, 
\textit{J. Phys. A: Math. Gen.} 
(to appear);
\texttt{nlin.SI/0103023}.

\bibitem{KM1} B. Konopelchenko and L. Martinez Alonso,
\textit{Phys. Lett. A}
\textbf{286}, 
161--166 (2001).

\bibitem{KM2} B. Konopelchenko and L. Martinez Alonso,
\textit{J. Math. Phys.}
(submitted); \texttt{nlin.SI/0105071}.

\bibitem{Vekua} I.N. Vekua, \textit{Generalized analytic functions},
Pergamon Press, Oxford (1962).

\bibitem{Kost} I.K. Kostov et al, 
{$\tau$-function for analytic curves},
in:
\textit{Random matrices and their applications, MSRI Publications},
v. \textbf{40}, 1--15 (2001).
 

\end{thebibliography}
\end{document}